\documentclass[]{spie}  %>>> use for US letter paper
\usepackage{amsmath,amsfonts,amssymb}
\usepackage{graphicx}
\usepackage[colorlinks=true, allcolors=blue]{hyperref}
\usepackage{float}
\usepackage{booktabs}
\usepackage{multirow}

\title{CCAT: Optical Design of the 410 GHz Prime-Cam Module}

\author[a]{Tilak M. Patel}
\author[a]{Tanner Buck}
\author[b]{Anthony Huber}
\author[a,d]{Eve M. Vavagiakis}
\author[b]{James Burgoyne}
\author[c,b]{Scott Chapman}
\author[d]{Ben Keller}
\author[a]{Jenna E. Moore}
\author[d,e]{Michael D. Niemack}
\author[d]{CCAT Collaboration}
\affil[a]{Department of Physics, Duke University, Durham, NC, USA}
\affil[b]{Department of Physics and Astronomy, University of British Columbia, Vancouver, Canada}
\affil[c]{Department of Physics and Atmospheric Science, Dalhousie University, Halifax, Canada}
\affil[d]{Department of Physics, Cornell University, Ithaca, USA}
\affil[e]{Department of Astronomy, Cornell University, Ithaca, USA}

\authorinfo{Further author information: Send correspondence to T.M.P.: E-mail: tilak.patel@duke.edu}

\begin{document} 

\maketitle

\begin{abstract}
Prime-Cam is a first-generation instrument for the Fred Young Submillimeter Telescope (FYST), enabling wide-field, multi-frequency observations for cosmology, line-intensity mapping, and galaxy studies. We present an optical performance study for a candidate 410 GHz broadband module designed to field approximately 21,000 polarisation-sensitive kinetic inductance detectors (KIDs). The three-lens silicon design was adapted from the SO LATR design and used for the existing 280 and 350 GHz Prime-Cam instrument modules, as well as this study. At 410 GHz, a shorter wavelength places tighter demands on wavefront quality and beam shape. Using Ansys Zemax OpticStudio and Huygens PSF analysis, we evaluate candidate module positions, compare 350 and 410 GHz performance, and assess field-dependent Strehl ratio, ellipticity, and encircled-energy behaviour. A preliminary tolerancing study, using inverse increment and Monte Carlo methods, tests sensitivity to selected alignment perturbations.
\end{abstract}


\section{INTRODUCTION}
\label{sec:intro}
\begin{figure}
    \centering
    \includegraphics[width=0.6\linewidth]{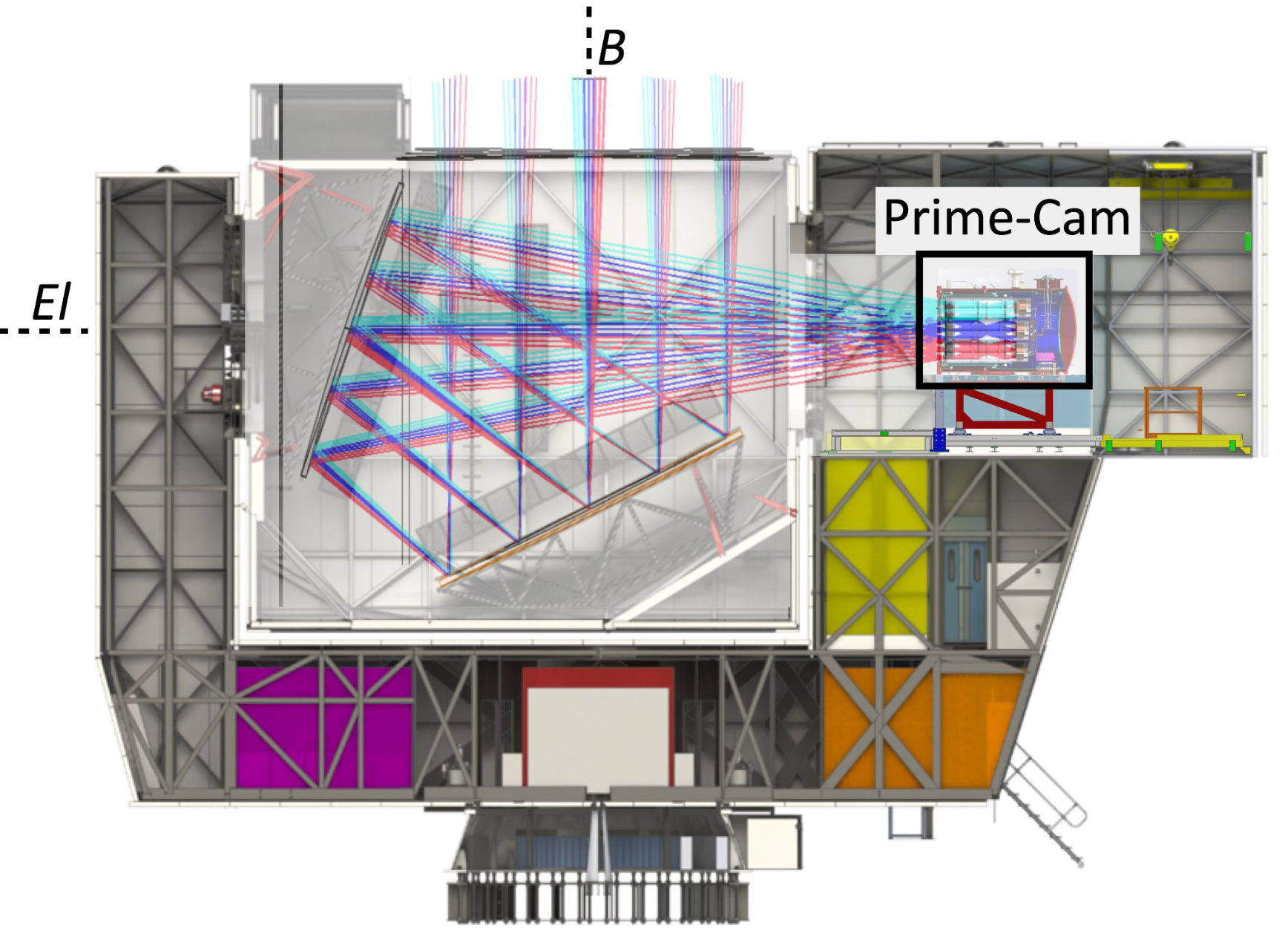}
    \caption{Prime-Cam in FYST, with ray traces showing the path of light guided into the instrument by two mirrors.}
    \label{fig:primecam}
\end{figure}

The Fred Young Submillimeter Telescope (FYST), shown in Fig. \ref{fig:primecam}\cite{vavagiakisPrimeCamInstrumentOverview2026}, is a 6-m crossed-Dragone telescope\cite{niemackDesignsLargeApertureTelescope2016,parshleyOpticalDesignSixmeter2018} designed for wide-field observations at mm and sub-mm wavelengths\footnote{\url{https://www.ccatobservatory.org}}. Prime-Cam is a first-generation instrument which will support several independent broadband imaging and spectroscopic instrument modules within a common cryogenic platform. \cite{vavagiakisPrimeCamFirstlightInstrument2018,huberCCATPrimeCamOptics2024,ccat-primecollaborationCCATprimeCollaborationScience2023,choiSensitivityPrimeCamInstrument2020}. A candidate broadband module centered at 410 GHz is partially funded, and when fully populated, will field approximately 21,000 polarisation-sensitive Kinetic Inductance Detectors (KIDs) \cite{Chapman2026}. Operation at 410 GHz places tighter demands on the optical system than the adjacent 350 GHz band\cite{kellerCCATDesignCharacterization2026}. For a fixed physical wavefront error, the corresponding phase error is larger with smaller wavelength\cite{wyantBasicWavefrontAberration1992}. Therefore, an optical design that performs comfortably at 350 GHz may exhibit lower Strehl ratio (lower peak intensity), greater field dependence, or more structured point spread functions (PSFs) when evaluated at 410 GHz. An existing optical architecture that was adapted from the longer wavelength SO LATR design \cite{dickerColdOpticalDesign2018,gallardoSystematicUncertainties2018} has already been developed for a Prime-Cam 350 GHz module\cite{huberConstrainingTimeDependentParity2025,kellerCCATDesignCharacterization2026}. Rather than beginning with an unconstrained redesign for 410 GHz, this analysis answers whether that mechanically compatible architecture remains viable at the shorter wavelength. The design is evaluated without changing any lens prescriptions or receiver geometries, so it provides a test of how far the established design can be carried before a more substantial redesign is necessary. 

Instrument module optical performance is not purely dependent on frequency, but also on the effective location of the module within Prime-Cam's focal plane. Different off-axis positions receive different aberrations due to the telescope's crossed-Dragone design. The nominal design is therefore tested for performance across available telescope configurations. Previous Prime-Cam optical studies have primarily summarised this position dependence using Strehl ratio\cite{huberConstrainingTimeDependentParity2025,huberCCATPrimeCamOptics2024,vavagiakisPrimeCamFirstlightInstrument2018,huberCCATprimeOpticalCryogenic2022}. Strehl provides a useful first-pass measure of peak image quality, but it does not uniquely describe the beam morphology. This analysis combines Strehl with Huygens point spread functions, generated in Zemax OpticStudio. The exported point spread functions are processed to calculate beam ellipticity, axis widths and orientations, full-width half-maximums (FWHMs), and encircled energy radii. The field sampling is also revised from inherited designs to better represent the detector populated region.

\section{Prime-Cam 410 GHz module optical design}
\label{sec:primecamdesign}

\subsection{Prime-Cam module architecture}
Prime-Cam is a 1.8 m diameter cryostat capable of accommodating up to seven independent instrument modules: one central module and six off-axis modules distributed around it\cite{vavagiakisPrimeCamFirstlightInstrument2018,huberCCATPrimeCamOptics2024}. Each module contains a dedicated set of cryogenic reimaging optics, filters, a Lyot stop, and detector arrays optimised for the specific observing band(s) and science goals\cite{huberCCATPrimeCamOptics2024, choiSensitivityPrimeCamInstrument2020}. Prime-Cam's modular architecture allows for multiple optical designs to coexist with a shared focus. Prime-Cam accommodates broadband imaging modules, spectrometer modules, and high-frequency cameras all within the same receiver, sharing common cryogenic and mechanical infrastructure \cite{vavagiakisCCATprimeDesignModCam2022, huberCCATprimeOpticalCryogenic2022,huberCCATPrimeCamOptics2024}.

Optical performance is not uniform across the Prime-Cam focal plane. While the central module position provides the best image quality, the off-axis positions experience varying levels of telescope-induced aberrations. Previous Prime-Cam optical studies therefore evaluated performance separately for each candidate module location using focal plane Strehl maps and detector array footprints \cite{huberCCATPrimeCamOptics2024, huberCCATprimeOpticalCryogenic2022}. The 850 GHz module is planned for the central position due to its more demanding optical requirements, whereas lower frequency modules can occupy outer locations\cite{huberOpticalMechanicalDetector2024,chapmanCCATprime850GHzCamera2022}. We must therefore assess candidate locations for a 410 GHz module in the context of both the internal module optics, and the position-dependent aberrations introduced by the telescope.

\subsection{Baseline 350 GHz three-lens design}
\begin{figure}
    \centering
    \includegraphics[width=0.6\linewidth]{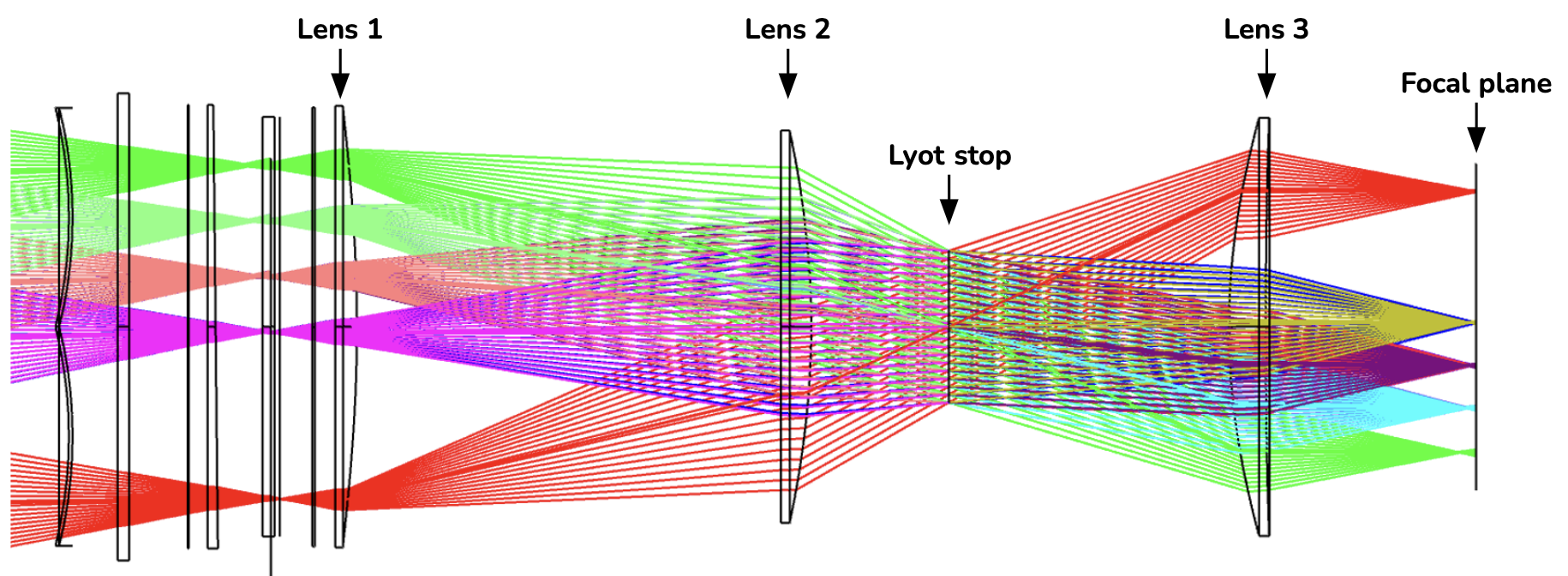}
    \caption{3-lens optical design ray trace generated by Zemax. The three lenses, focal plane, and lyot stop are labelled.}
    \label{fig:raytrace}
\end{figure}

The 410 GHz optical design we analyse is based on the three-lens silicon architecture used in Prime-Cam's 350 GHz instrument module, which inherited the 280 GHz instrument module design \cite{kellerCCATDesignCharacterization2026,vavagiakisCCATprimeDesignModCam2022,huberCCATPrimeCamOptics2024}. Fig.~\ref{fig:raytrace} displays a ray-trace diagram of the design. This choice is driven by both practical and optical considerations. Repurposing an existing optical design allows lens manufacturers to avoid fabrication complexities that come with designing new custom lenses. It also avoids introducing additional fabrication requirements and new alignment degrees of freedom.

The 350 GHz instrument module provides a natural design reference point as the target frequencies are relatively similar and the three-lens design reduces fabrication cost. The main concern is that the shorter wavelength of the 410 GHz module makes the same physical wavefront errors more significant. A design that is near diffraction-limited at 350 GHz may therefore exhibit lower Strehl, increased field dependence, or altered beam morphology when evaluated at 410 GHz. The purpose of this paper is to assess how well this existing mechanically compatible architecture performs at our target frequency and how far it can be pushed before performance falls below a satisfactory level.

The value of starting from this design is that it keeps the optical analysis tied to an existing Prime-Cam instrument module. The initial constraint driving the design is the mechanical envelope of Prime-Cam which restricts the volume and dimensions of each module. The existing three lens layout meets this requirement, as well as the specifications for optical path length, detector footprint, and cryogenic interfaces. 

\subsection{Candidate Prime-Cam module positions}
\begin{figure}
    \centering
    \includegraphics[width=1.08\linewidth]{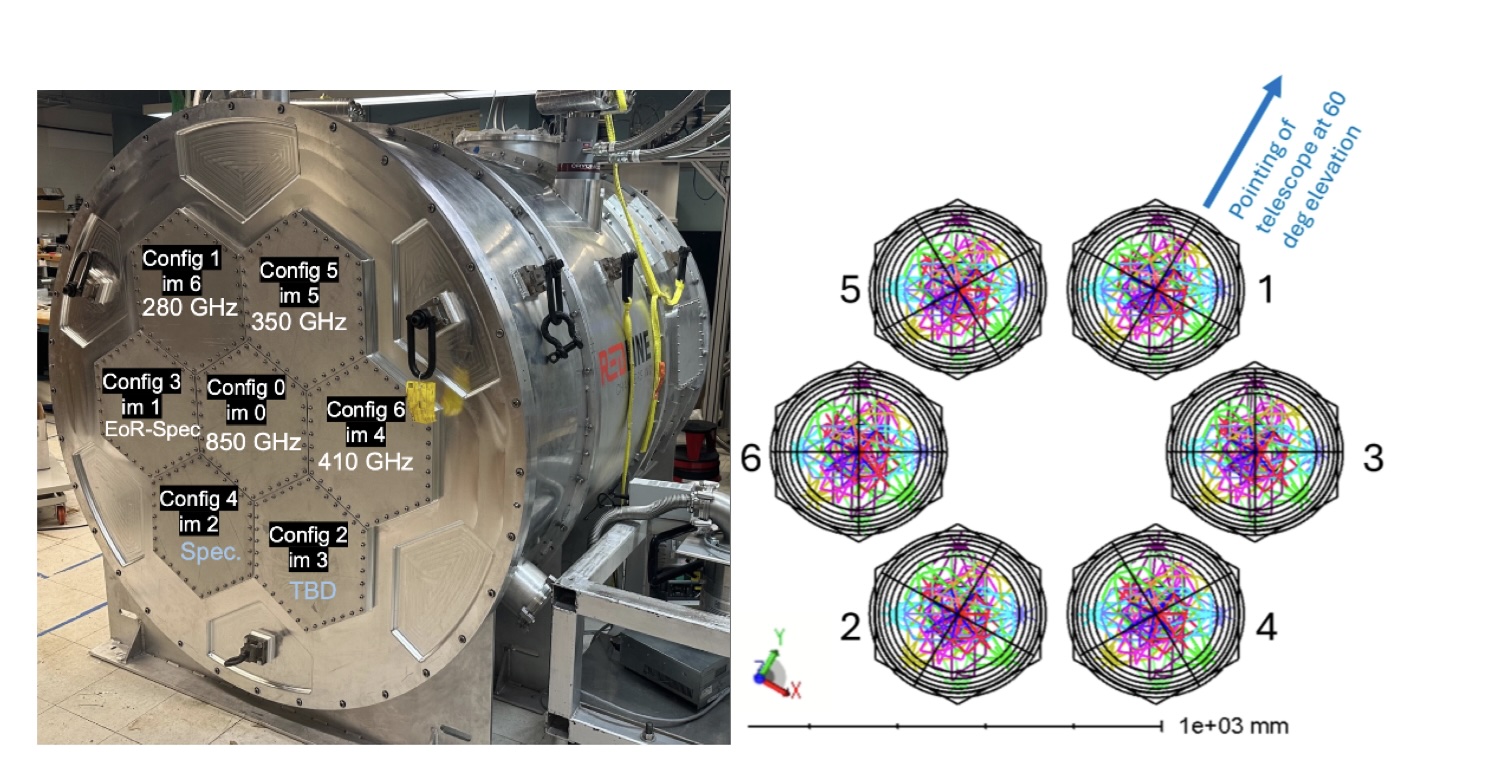}
    \caption{Photo of the Prime-Cam instrument showing the instrument module positions and the relevant optical tube/configuration positions considered for the 410 GHz module. The right plot is a Zemax simulation viewed from the back.}
    \label{fig:placeholder}
\end{figure}

Prime-Cam is designed to house seven instrument modules, across one central configuration, Config 0, and six off axis configurations, Configs 1-6 (See Fig.~\ref{fig:placeholder}). This \textit{Config} naming convention is specific to optical simulations - mechanically, the optical tubes are labelled with an instrument module-standard. The optical performance of a module depends, in part, on its location in the receiver's focal plane.  Although each module is designed to uniquely reimage the light passing through for its specified frequency, the incoming beam has already passed through the telescope. Hence, different positions experience different telescope-induced aberrations before the optical components are introduced. A design that performs well in one position is not guaranteed to perform similarly in other positions.

We evaluate the candidate 410 GHz module in several effective Prime-Cam optical configurations, corresponding to Zemax configurations 2, 3, 4, and 6, which represent available locations in Prime-Cam after accounting for established and planned modules. These describe the optical state seen by a module at different locations in the outer Prime-Cam focal plane. It is important to make the distinction between physical module location and effective optical configuration. The physical location is constrained by the already occupied locations. 850 GHz is planned for the central Config 0, due to its shortest wavelength and hence highest sensitivity demanding the most forgiving location in Prime-Cam. The best performing off-axis configuration is known to be configuration 3, which is reserved for the Epoch-of-Reionization Spectrometer (EoR-Spec), Prime-Cam's line intensity mapping instrument\cite{freundtStatusUpdateEoRSpec2024,nikolaEpochReionizationSpectrometer2023}. However, because the telescope pointing changes how the receiver samples the focal plane, observations can be planned so that the 410 GHz module operates with an effective optical configuration equivalent to Config 3. This behaviour is comparable to a Nasmyth-like focus, where the instrument remains fixed while the orientation of the incoming beam changes with telescope elevation. As such, we include this configuration in our analysis, even though the module may not physically occupy the slot in Prime-Cam\cite{huberConstrainingTimeDependentParity2025}. The goal is to determine the best effective optical configuration for optical performance for the 410 GHz design. 

The three lens optical design is kept constant across each configuration, to ensure the comparison is focused on the optical consequences of effective module location rather than changes to the internal design. Differences in beam characteristics can therefore be attributed directly to module position.

\section{Simulation and analysis methods}
\label{sec:simulationanalysismethods}

\subsection{Zemax optical model}
The optical simulations were carried out in \textit{Ansys Zemax OpticStudio}\footnote{\url{https://www.ansys.com/products/optics/ansys-zemax-opticstudio}}, a ray tracing and optical design software widely used for sequential optical systems. OpticStudio is a software that enables the modelling of the telescope and receiver optics in a single prescription, the support of multi-configuration analyses for comparing different optical states, and its built-in tools for optimisation, image-quality analysis, physical PSF calculations, and tolerancing features. This made it possible to preserve the existing Prime-Cam optical model while changing targeted parameters such as wavelength and effective module configuration, without rebuilding the full system from scratch. 

The Zemax model used in this work was based on the Prime-Cam 350 GHz broadband module optical system\cite{kellerCCATDesignCharacterization2026}. This archive already contained the FYST crossed-Dragone telescope prescription, the Prime-Cam receiver coordinate system, the three lens module optics, and the multi-configuration structure used to represent the six off-axis positions in Prime-Cam. The model is organised in the multi-configuration editor (MCE). Each configuration corresponds to a different effective optical location in Prime-Cam. All simulations presented in this work were performed using the inherited Zemax model at a fixed telescope elevation of $60^{\circ}$. Changing telescope elevation rotates the effective Prime-Cam configurations about the receiver axis. The analysis therefore compares candidate configurations under a common reference telescope orientation. Each configuration places the module at a different location on the telescope focal plane, causing it to sample a different part of the telescope-induced aberrations before the beam enters the module optics.

To evaluate the 410 GHz optics, the wavelength was changed directly in the Zemax wavelength editor to 730\,$\mu$m. The lens surfaces, spacings, MCE structure, and receiver geometry were otherwise left unchanged. The 350\,GHz system was retained as the reference model, with the modified design being treated as a direct test of the same optical architecture at the shorter wavelength. This provides a controlled comparison, with differences in performance being purely attributable to the difference in wavelength. For each selected configuration and frequency, Zemax was used to generate field-dependent Strehl information and Huygens point spread functions. The Huygens PSFs were exported as text files and post-processed in Python.

\subsection{Huygens PSF analysis}
The PSF calculation was used because it provides a field-dependent intensity distribution, rather than reducing the optical performance to a single scalar quantity. Previous Prime-Cam optical analyses have often used Strehl ratio as the main way to assess optical performance\cite{huberCCATPrimeCamOptics2024,huberCCATprimeOpticalCryogenic2022,huberCCATprimeOpticalDesign2022}. This is a natural first metric, as it provides a simple scalar that quantifies image quality. However, it does not contain all information needed to judge beam quality. Two beams with similar Strehl ratios can differ in beam width, ellipticity, orientation, etc. Huygens PSFs give the two dimensional intensity distribution in the image plane at a specific point\cite{ansysHuygensPSF}. Therefore, each beam is assessed by its peak image quality, as well as its shape and compactness.

Huygens PSFs were exported for configurations 2, 3, 4, and 6 at both 350 GHz and 410 GHz. For each configuration and frequency combination, the PSF was computed at each of the 25 field points. These were then processed in Python, with the intensity array for each field point being paired with the corresponding image-plane coordinate grid, and the beam properties computed directly from the sampled PSF. The same process was used for each configuration and frequency combination, ensuring consistency across the analysis.

\subsection{Optical performance metrics}
The optical performance of each configuration was quantified using a set of PSF-derived metrics, presenting a holistic approach to optical performance analysis in comparison to the approach performed in prior Prime-Cam studies. These metrics separate different aspects of a beam: peak intensity, beam width, beam symmetry, and the concentration of power within the beam. This is useful because no single metric can fully describe the optical quality of the module; Strehl ratio is only sensitive to peak degradation, whilst other metrics tell us how the beam is physically distributed\cite{mahajanStrehlRatioPrimary1983}.

The primary diffraction-limited metric used in this work is the Strehl ratio. It is defined as the ratio of the peak intensity of the aberrated PSF to the peak intensity of an ideal diffraction-limited PSF for the same optical system. Diffraction-limited is defined as a Strehl greater than 0.8. 

The width of the main PSF beam was quantified using the FWHM. This is the width of the beam at half of its peak intensity. The FWHM is taken in both the x and y directions due to the fact that PSFs are rarely perfectly circular. These values provide a simple measure of the beam size, but do not fully describe asymmetric or rotated beams, as opposed to ellipticity which is a good measure of symmetry. Ellipticity is calculated as the relative difference between the major and minor second moment widths, with low ellipticity indicating a nearly circular beam, and therefore better optical performance: the larger the ellipticity, the more asymmetric the beam is, causing distorted image quality. 

To measure how compactly the power in the PSF is concentrated, we also considered the encircled-energy radii. The quantity $r_{80}$ gives the radius enclosing 80\% of the PSF power, while $r_{95}$ is the same quantity for 95\% of the power. In general, $r_{80}$ is used as the main metric to describe compactness, while $r_{95}$ is used as an additional measure of the extended beam structure. As such, this analysis will focus mainly on $r_{80}$. These quantities take into account the power distributed outside the central peak in secondary lobes.

Together, Strehl, ellipticity, and encircled energy radius, provide a more complete description of optical performance than Strehl ratio alone. They allow us to distinguish between different forms of beam degradation, including reduced peak intensity, broadening of the main lobe, asymmetry of the beam, and power distribution.

\subsection{Detector-aware field sampling}
\begin{figure}[H]
    \centering
    \includegraphics[width=0.5\linewidth]{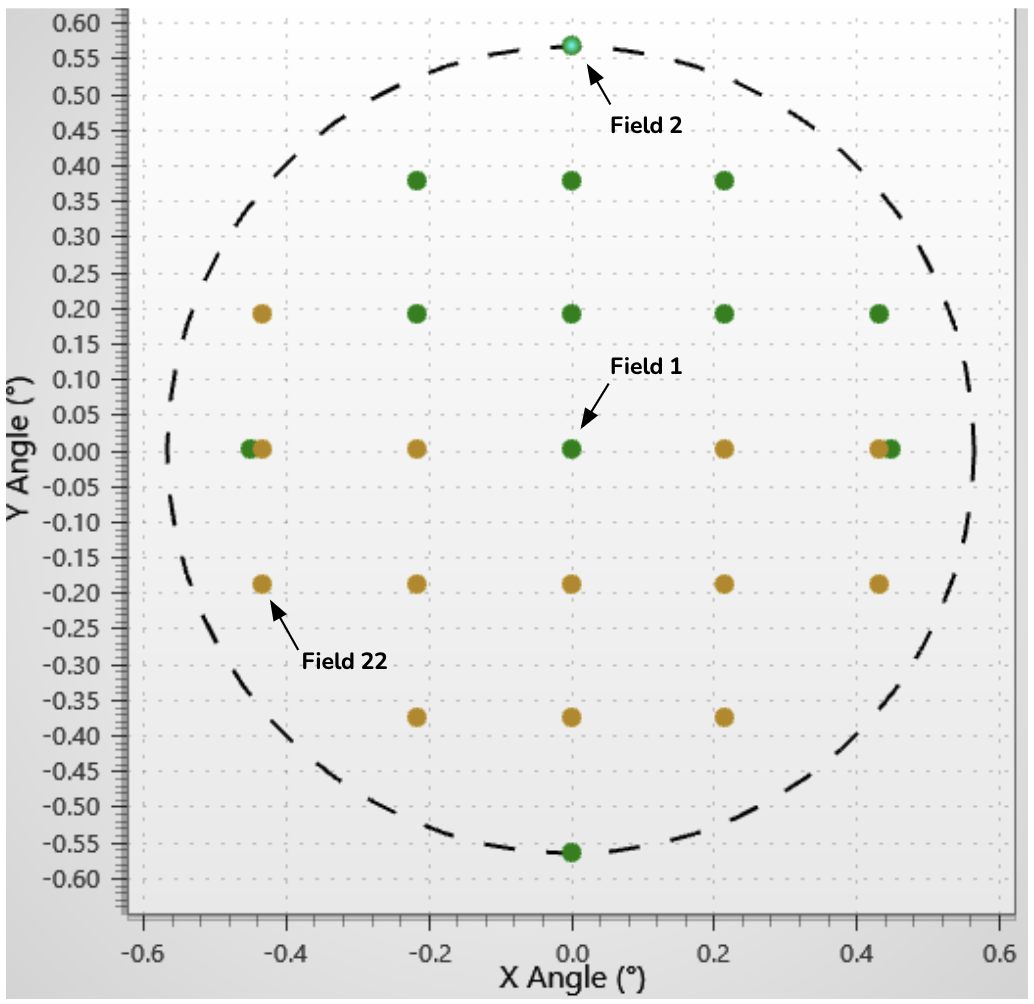}
    \caption{Detector-aware field sampling used for the analysis. The 25 field points are defined in angular coordinates and assigned equal weight in the calculated averages. Fields 1, 2, and 22 are labelled.}
    \label{fig:updatedfieldset}
\end{figure}

The optical performance was evaluated at 25 discrete field points across the focal plane (Fig.~\ref{fig:updatedfieldset}). These field points represent angular sample locations at which Zemax computes the PSF. Relative to the field set used by Ref.~\citenum{huberConstrainingTimeDependentParity2025}, fields 4 and 5 were moved from x = $\pm0.65^\circ$ to x = $\pm0.45^\circ$ to better represent the populated detector footprint while retaining off-axis sampling. This modification prevents two extreme edge samples from contributing equally to the field-averaged performance metrics.

Table~\ref{tab:detector_aware_sampling} summarises the effect of the updated field sampling for configuration 3. Moving the two outermost field points inward primarily improves the minimum Strehl ratio and maximum ellipticity, while producing only modest changes in the field-average metrics. The revised field set was therefore adopted across all configurations and frequencies throughout the remainder of the analysis.

\begin{table*}[t]
\centering
\caption{Effect of updating fields 4 and 5 for Config 3. The original field set placed fields 4 and 5 at $x=\pm0.65^\circ$, while the detector-aware field set moved them inward to $x=\pm0.45^\circ$ to better represent the detector footprint. Strehl ratio and ellipticity are dimensionless, while $r_{80}$ and $r_{95}$ are encircled energy radii measured in the image plane.}
\label{tab:detector_aware_sampling}

\begin{tabular}{llcccccc}
\toprule
\multirow{2}{*}{\textbf{Frequency}} &
\multirow{2}{*}{\textbf{Sampling}} &
\multicolumn{2}{c}{\textbf{Strehl ratio}} &
\textbf{Maximum} &
\multicolumn{2}{c}{\textbf{Encircled-energy radius ($\mu$m)}} \\
\cmidrule(lr){3-4}
\cmidrule(lr){6-7}
&
&
\textbf{Mean}
&
\textbf{Minimum}
&
\textbf{Ellipticity}
&
$\mathbf{r_{80}}$
&
$\mathbf{r_{95}}$
\\
\midrule

\multirow{2}{*}{350 GHz}
& Original
& 0.866
& 0.579
& 0.366
& 2071
& 5054
\\

& Detector-aware
& \textbf{0.879}
& \textbf{0.723}
& \textbf{0.104}
& \textbf{1991}
& \textbf{4986}
\\

\midrule

\multirow{2}{*}{410 GHz}
& Original
& 0.822
& 0.468
& 0.428
& 1920
& 4339
\\

& Detector-aware
& \textbf{0.838}
& \textbf{0.638}
& \textbf{0.105}
& \textbf{1844}
& \textbf{4270}
\\

\bottomrule
\end{tabular}
\end{table*}

\section{Configuration comparison}
\label{sec:configcomparison}

\subsection{Candidate-position performance at 410 GHz}
\begin{figure}[H]
    \centering
    \includegraphics[width=\linewidth]{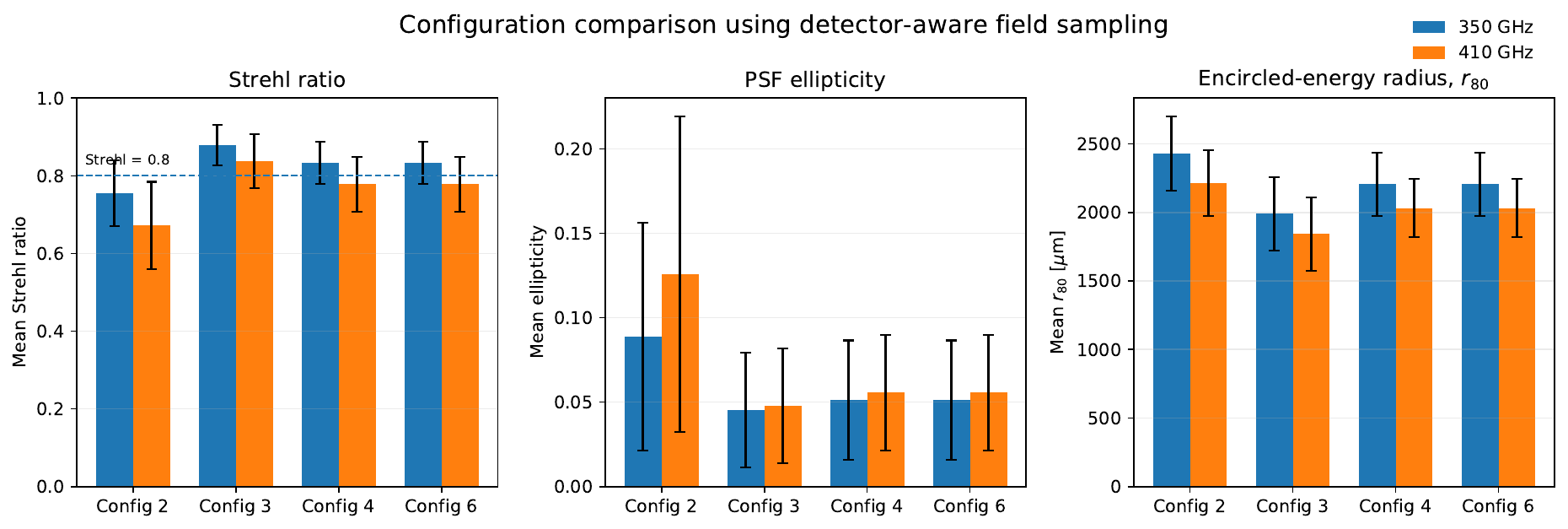}
    \caption{Comparison of performance at effective Prime-Cam configurations for 350 GHz and 410 GHz modules. Bars show the mean across the 25 field samples, while error bars show the field-to-field standard deviation. The dashed line (S=0.8) marks the Strehl threshold. Configuration 3 gives the strongest combined performance, with the highest mean Strehl ratio, lowest mean ellipticity, and smallest mean \(r_{80}\).}
\label{fig:config_comparison}
\end{figure}

We first compare the optical performance at each of the four effective configurations for the 410 GHz module (Configs 2, 3, 4, and 6). The resulting performance metrics are summarised in Table~\ref{tab:config_summary} and Fig.~\ref{fig:config_comparison}. The mean values describe the performance across each configuration, while the minimum Strehl, maximum ellipticity, and standard deviations indicate how much the PSF quality varies field to field. This distinction is important since a configuration can have an acceptable average while still containing localised regions with degraded beams. 

At 410 GHz, configuration 3 is the strongest of the four candidate configurations. It has a mean Strehl ratio of 0.838, which is the only location that has a mean value above the 0.8 threshold. It also has the lowest mean ellipticity, 0.048, and the smallest mean $r_{80}$, 1844 $\mu$m. This position exhibits the best combined result in peak image quality, beam symmetry, and encircled-energy compactness. On the contrary, configuration 2 is clearly the weakest effective position. At 410 GHz, its mean Strehl ratio is 0.672, far below the threshold, while its beam shape metrics are also worse than the other configurations. Mean ellipticity is 0.126, maximum ellipticity is as high as 0.377, and mean $r_{80}$ is 2212 $\mu$m. Compared with configuration 3, configuration 2's mean Strehl is 20\% lower, its mean ellipticity is more than double, and it has a broader $r_{80}$ by nearly 370 $\mu$m.

The field-to-field scatter reinforces this conclusion. Configuration 2 has a Strehl standard deviation of 0.112, with an ellipticity standard deviation of 0.093, both of which are larger than those for configuration 3. This means that configuration 2 isn't only worse on average, but it is also less uniform across the sampled field. Some regions of the field are especially degraded, which is why the maximum ellipticity is so large. This behaviour reinforces the necessity of analysing PSF metrics, as mean Strehl alone would understate these effects.

Configurations 4 and 6 form an intermediate pair, with identical summary values. They have a mean Strehl of 0.778, mean ellipticity of 0.056, and mean $r_{80}$ of 2031 $\mu$m. They perform more similarly to configuration 3 than configuration 2, so they serve as plausible backup configurations, but not the preferred solution as they are still subordinate to configuration 3.

The configuration choice is therefore unambiguous. For the fixed three lens design we evaluated, the 410 GHz module performs best when placed at an effective configuration 3 optical state. Configuration 2 is unfavourable due to its poor performance. Configurations 4 and 6 have similar performance, but under-perform relative to configuration 3. This motivates using configuration 3 as the baseline case for the direct frequency comparison. 

\subsection{Comparison between 350 GHz and 410 GHz}
The 350 GHz optical simulation provides the reference case for studying its performance at 410 GHz. For configuration 3, the mean Strehl decreases from 0.879 at 350 GHz to 0.838 at 410 GHz. This is a moderate decrease of $\sim 4.7\%$, but importantly, it remains above the threshold of 0.8, which means that configuration 3 remains diffraction-limited on average across the field of view. The minimum Strehl does decrease from 0.723 to 0.638, showing that the frequency penalty is more pronounced in the weaker parts of the sampled field, than in the mean alone. The beam shape metrics change much less. The mean ellipticity increases modestly by $\sim 6.7\%$, from 0.045 to 0.048 at 410 GHz. The max ellipticity is also essentially unchanged, from 0.104 to 0.105. This is a key result as it shows that the penalty is not accompanied by a major increase in beam asymmetry for configuration 3. The encircled energy behaviour also remains controlled. The mean $r_{80}$ decreases from 1991 $\mu$m to 1844 $\mu$m at 410 GHz. At first, this appears to have improved, as it represents a more compact beam. However, because we are looking at a beam with a shorter wavelength here, the actual beam size decreases, so a lower $r_{80}$ value is expected. The key finding is that the PSF remains compact and the ellipticity does not grow substantially. The other configurations scale in similar ways.

Configuration 2 exhibits a much less favourable frequency response. Its mean Strehl decreases from 0.755 at 350 GHz to 0.672 at 410 GHz, representing a more substantial 11\% decrease. Its mean ellipticity also increases from 0.089 to 0.126, and the max ellipticity rises from 0.293 to 0.377. This means that configuration 2 is suffering both a high Strehl penalty, and substantially worse beam asymmetry across parts of the field. Configurations 4 and 6 again display intermediate performance, with mean Strehl decreasing 6.7\% from 0.834 at 350 GHz to 0.778 at 410 GHz. Their mean ellipticity increases from 0.051 to 0.056, and mean $r_{80}$ decreases from 2205 $\mu$m to 2031 $\mu$m. These changes are more moderate than configuration 2 but still less favourable than configuration 3. In particular, the mean Strehl falls below the diffraction-limited threshold value of 0.8, whereas configuration 3 remains above it. 

The 350 GHz to 410 GHz comparison therefore supports the main configuration result. The three lens architecture does exhibit degraded performance at the higher 410 GHz frequency, as expected, but the degradation is not catastrophic in the preferred configuration 3 position. In configuration 3, the dominant penalty is a moderate Strehl reduction, while ellipticity remains low and the encircled-energy radius remains compact. Configuration 2 is disfavoured at both frequencies and becomes substantially worse at 410 GHz. Configurations 4 and 6 remain usable comparison cases, but they do not match the performance of configuration 3. Overall, the results indicate that the 350 GHz design is viable at 410 GHz if the effective optical configuration is chosen carefully, with configuration 3 providing the best current baseline for further field resolved PSF analysis and tolerance testing. 

\begin{table}[H]
\centering
\small
\caption{Summary of PSF-derived optical performance metrics for candidate effective Prime-Cam configurations at 350 and 410 GHz. Metrics are computed using the field sampling from Fig.~\ref{fig:updatedfieldset}. Strehl ratio and ellipticity are dimensionless. \(r_{80}\) and \(r_{95}\) are encircled-energy radii.}
\begin{tabular}{c c c c c c c c}
\hline
Freq. & Config &
Strehl & Strehl$_{\min}$ &
Ellipticity &
$\overline{r}_{80}$ & $\overline{r}_{95}$ \\
GHz & & &
 &
 &
$\mu$m & $\mu$m \\
\hline
350 & 2 & 0.755 & 0.655 & 0.089 & 2428 & 4992 \\
350 & 3 & 0.879 & 0.723 & 0.045 & 1991 & 4986 \\
350 & 4 & 0.834 & 0.715 & 0.051 & 2205 & 4968 \\
350 & 6 & 0.834 & 0.715 & 0.051 & 2205 & 4968 \\
\hline
410 & 2 & 0.672 & 0.554 & 0.126 & 2212 & 4294 \\
410 & 3 & 0.838 & 0.638 & 0.048 & 1844 & 4270 \\
410 & 4 & 0.778 & 0.629 & 0.056 & 2031 & 4256 \\
410 & 6 & 0.778 & 0.629 & 0.056 & 2031 & 4256 \\
\hline
\end{tabular}
\label{tab:config_summary}
\end{table}

\section{Field-resolved PSF Performance}
\label{sec:psfmorph}
The configuration-level results in Section~\ref{sec:configcomparison} identify configuration 3 as the preferred effective position for the candidate 410 GHz module. Across the detector-aware field set, its PSF metrics show consistently better performance. These statistics establish a clear field averaged preference, but they do not describe the form of the underlying beams. A PSF can have a high normalised peak while displaying asymmetry, or remain almost circular but have power transferred out of the central core. The metrics are complementary, because none of them provide a complete description of the beam on their own. We examine the Huygens PSFs here to determine how different forms of degradation appear in the image plane. The comparison is restricted to configurations 2 and 3 at 410 GHz, as the former displays the weakest aggregate performance, while the latter shows the strongest. Because configurations 4 and 6 just display intermediate performance and are nearly identical, analysis of these will not provide as much insight. Focusing on the bounding cases makes the physical origin of the configuration ranking easier to interpret.

\subsection{Matched field Huygens PSFs}
\begin{figure}[H]
    \centering
    \includegraphics[width=0.8\linewidth]{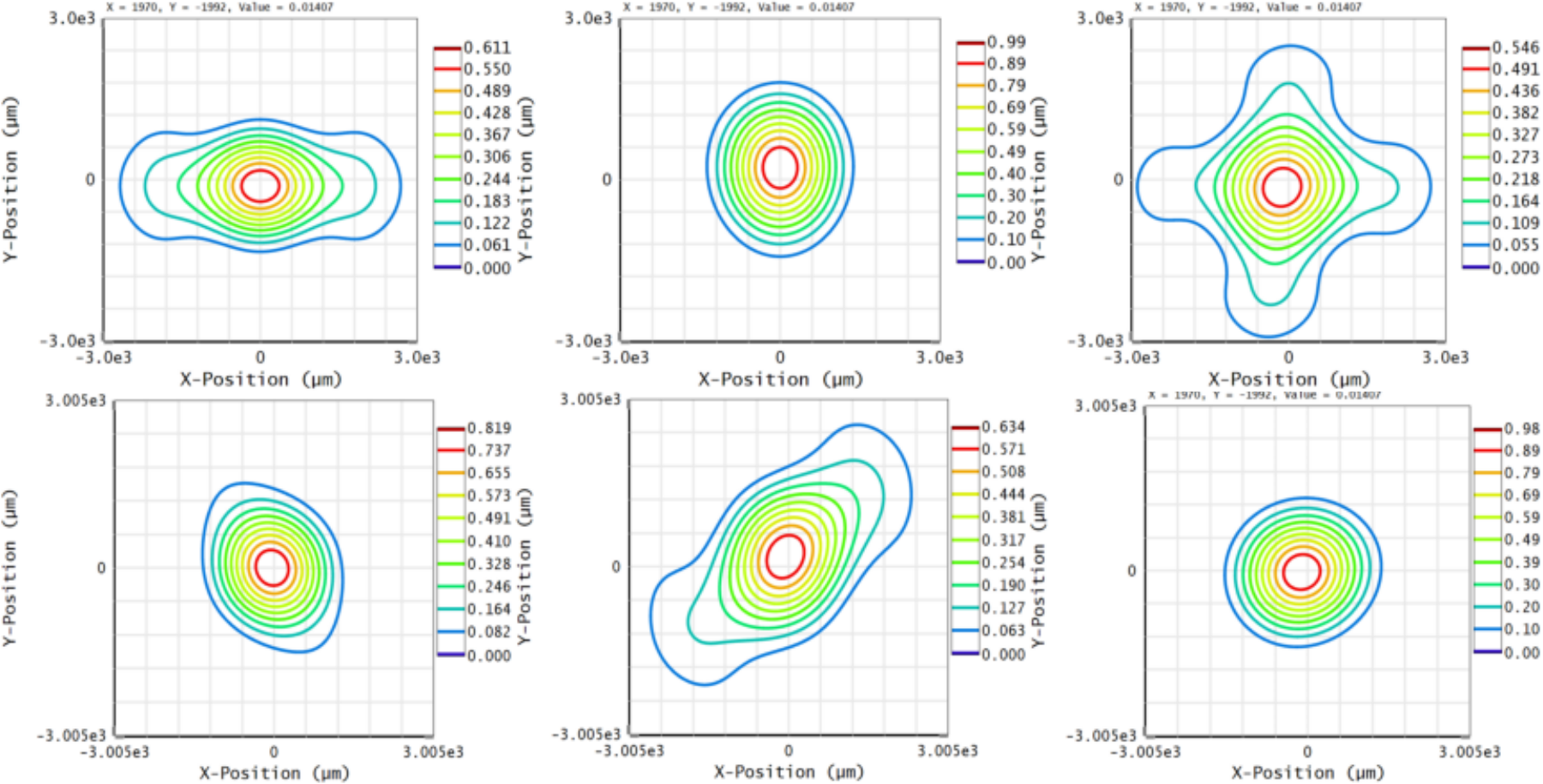}
    \caption{Matched 410 GHz Huygens PSFs for effective configurations 2 (top row) and 3 (bottom row). Each column compares the same angular field coordinate. Field 1 (left) provides the central reference, Field 2 (middle) lies at the positive y edge of the field, and Field 22 (right) lies in the negative x, negative y portion of the field. All PSFs were generated using identical image plane dimensions, sampling, centroid normalisation, and contour levels. The comparison illustrates that Strehl ratio, ellipticity, and encircled energy radius do not necessarily rank the beam in the same order, as described in Table~\ref{tab:matched_psf_fields}.}
    \label{fig:matched_psfs}
\end{figure}

Figure~\ref{fig:matched_psfs} shows the normalised Huygens PSFs at three field points intended to represent the field of view: Fields 1, 2, and 22. Configuration 2 is shown in the upper row and configuration 3 in the lower row. Each column corresponds to the same angular field positions in both configurations.

Field 1 provides the central reference. Field 2 lies at the positive-y boundary of the field, at (x,y) = (0,0.566$^{\circ}$). Field 22 lies in the negative-x, negative-y portion of the field, at (x,y) = (-0.433$^{\circ}$,-0.189$^{\circ}$). We selected fields 2 and 22 because they undergo opposite changes in Strehl performance between the configurations. Field 2 has the highest Strehl in configuration 2 and lowest in configuration 3, whereas Field 22 changes from the lowest Strehl in configuration 2 to the highest in configuration 3. 

The central PSFs are compact in both configurations, but are not identical. The outer configuration 2 contours are extended predominantly along the horizontal direction, producing a central beam with visible lateral shoulders. In configuration 3, the corresponding PSF is more compact, although the contours do display a small tilt and depart slightly from perfect circular symmetry. The Field 1 comparison therefore shows that configuration dependent morphology is present even near the centre of the sampled field, although the changes are less dramatic than at the two off-axis locations.

Field 2 provides the clearest example of a position that favours configuration 2. The PSF is compact, with fairly circular contours and only slight asymmetry. The Strehl ratio is 0.989 with an $r_{80}$ of 1472 $\mu$m, which is consistent with the strong central concentration shown in Figure~\ref{fig:matched_psfs}. Evaluating Field 2 in configuration 3 results in a considerably broader PSF with a more diagonal extension. The outer contours are displaced along a diagonal axis, although the inner contours are still fairly concentrated around the peak. Its Strehl ratio decreases to 0.638 and $r_{80}$ increases to 2559 $\mu$m. The change is not only limited to a reduction in the intensity of the peak, but the power is being distributed over a larger region of the plane. The ellipticity for Field 2 also drops from 0.150 in configuration 2 to 0.068 in configuration 3. At first glance, it seems that this is inconsistent with the visibly extended contours in configuration 3. This comparison illustrates an important limitation of reducing a PSF to just a single value quantifying asymmetry. Ellipticity quantifies the net imbalance between two widths, but it does not necessarily describe the lobes, especially when they are distributed on both sides of the centroid. Therefore, a low ellipticity value can coexist with a PSF that is visibly non-Gaussian and extended.

Field 22 reveals a more striking failure mode. In configuration 2, the PSF has a four-lobed outer contour, which produces a clover shape. Its Strehl ratio is 0.554 and $r_{80}$ is 2403 $\mu$m, indicating that the central peak is reduced strongly, and that a substantial fraction of the power is distributed away from the core. Despite the suboptimal morphology, the ellipticity is only 0.015. This is due to the four lobes being distributed in a balanced pattern, with the widths along each principal axis being nearly equal. The PSF is therefore almost circular in a second moment sense, even though it is not compact or close to diffraction-limited. Field 22 is therefore one of the clearest examples in the study of why low ellipticity cannot be interpreted as evidence of good overall beam quality. In configuration 3, there is no four-lobed structure for Field 22. The PSF becomes compact and approximately circular, with closely spaced contours surrounding the central maximum. Its Strehl ratio increases from 0.554 to 0.985, while $r_{80}$ decreases from 2403 to 1376 $\mu$m. The ellipticity rises slightly from 0.015 to 0.058, but this increase does not signify a degradation of the overall beam. The configuration 3 PSF is visibly more compact, and does not display the clover structure from configuration 2. The small increase in ellipticity is outweighed by an improvement in peak image quality and encircled-energy compactness.

The three matched comparisons show that changing effective configuration does not simply sharpen or broaden every field by a common amount. At Field 1, the beam changes fairly modestly. At Field 2, configuration 3 shows a broad, diagonally structured PSF where configuration 2 is compact. At Field 22, configuration 3 does not have the four-lobed structure and restores a compact central beam. Therefore, the incoming light's aberration pattern interacts with the optical design differently at each field location. This result also provides more information on the ranking of configurations from Section~\ref{sec:configcomparison}. While not superior in all metrics, configuration 3 is preferred statistically and across the detector-relevant field as a whole. Field 2 is an exception to this. The practical value of configuration 3 is in its more favourable overall distribution of beam quality, particularly its lack of severe morphology failures which are found in configuration 2.

\subsection{Beam morphology beyond scalar optical metrics}
The matched PSFs demonstrate why the optical system cannot be assessed using Strehl ratio alone. Strehl is an efficient and physically meaningful first pass measure of wavefront quality. It is also particularly useful for evaluating large field grids and for identifying where performance satisfies the diffraction-limited threshold. However, what Strehl does not provide is a description of how the power that is removed from the central peak is spatially arranged. The Field 2 and Field 22 PSFs illustrate two very different low Strehl behaviours. Configuration 3 Field 2 is broad and diagonally extended. Configuration 2 Field 22 has a more balanced four-lobed structure previously discussed. Both beams have a reduced peak image quality relative to their preferred configuration counterparts, but the power is distributed in different ways. A single Strehl scalar cannot distinguish between these morphologies.

Ellipticity adds insight to directional symmetry, but it also has limitations. It is most intuitive for a beam that can be approximated by a single elliptical central core. However, it becomes less insightful when there are multiple lobes, rings, and other structures. Configuration 2 Field 22 demonstrates this directly, with the PSF far from a compact diffraction-limited beam, yet it produces an ellipticity of only 0.015. 

The encircled-energy radii are a particularly useful metric in this case. Unlike ellipticity, $r_{80}$ does not require the redistributed power to favour a particular axis. Any broadening, lobes, etc. that carry a significant fraction of the total power increases the radius required to enclose 80\% of that power. The large configuration 2 Field 22 value of 2403 $\mu$m therefore captures an important aspect of the beam that is almost absent from its ellipticity. The comparison also shows why it is not trivial to label fields as `best' or `worst'. Field 2 is the best configuration 2 field when considering Strehl, but it is not the most circular beam. Field 22 is the worst configuration 2 field by Strehl, but its ellipticity is close to zero. A field can therefore rank first under one PSF metric and poorly under another. The matched PSFs strengthen the preference for configuration 3, but they also make that preference more precise. Configuration 3 is not a uniformly corrected configuration; it retains field dependence and includes at least one substantially degraded field point, in Field 2. Its advantage is that the field wide population is better controlled, with a higher mean Strehl, lower field-to-field scatter, and a smaller average encircled-energy radius. The Field 22 transformation provides the most visually compelling evidence for this conclusion. Configuration 3 converts an extended clover-like structure into a compact central beam, with a large gain in Strehl and reduction in $r_{80}$. Field 2 provides the necessary counterexample, showing that the configuration change can move the burden of the aberration rather than eliminate it. Together, the two fields explain why the final design decision must be based on the distribution of performance over the populated focal plane rather than just one favourable or unfavourable ray bundle.

Sections \ref{sec:configcomparison} and \ref{sec:psfmorph} have complementary roles in this analysis. The configuration comparison establishes which effective optical state is favoured statistically. The matched Huygens PSFs show the underlying beam structure, and reveal where the scalar metrics agree, where they diverge, and why all three are required. This combined evidence provides a stronger basis for carrying configuration 3 forward into the tolerance analysis than a Strehl-based argument alone. 

\begin{table}[H]
\centering
\small
\caption{PSF-derived metrics for the matched field positions shown in Fig.~\ref{fig:matched_psfs}. Field 1 provides the on-axis reference. Field 2 illustrates a position that favors Config 2 in Strehl ratio and encircled-energy compactness, while Field 22 illustrates the recovery of a compact, high-Strehl PSF in configuration 3.}
\begin{tabular}{c c c c c c c}
\hline
Field & Config & $x$ & $y$ & Strehl ratio & Ellipticity & $r_{80}$ \\
 & & (°) & (°) & & & ($\mu$m) \\
\hline
1  & 2 &  0.000 &  0.000 & 0.618 & 0.262 & 2333 \\
1  & 3 &  0.000 &  0.000 & 0.823 & 0.039 & 1946 \\
\addlinespace[2pt]
2  & 2 &  0.000 &  0.566 & 0.989 & 0.150 & 1472 \\
2  & 3 &  0.000 &  0.566 & 0.638 & 0.068 & 2559 \\
\addlinespace[2pt]
22 & 2 & -0.433 & -0.189 & 0.554 & 0.015 & 2403 \\
22 & 3 & -0.433 & -0.189 & 0.985 & 0.058 & 1376 \\
\hline
\end{tabular}
\label{tab:matched_psf_fields}
\end{table}

\section{Tolerance Sensitivity and Monte Carlo Robustness}
The nominal optical results establish that the 350 GHz three-lens design can provide useful 410 GHz performance in configuration 3. However, this does not establish that the design is practical. The fabricated system will have finite errors in lens placements and orientation, and several of those errors will be present simultaneously. A design that performs well at only its exact Zemax prescription places unrealistic demands on the mechanical structure and integration procedure. Therefore, we conducted a tolerancing analysis that was designed around two related questions. The first is diagnostic: which alignment degrees of freedom produce the largest change in optical performance when varied individually? The second is statistical: when a set of errors are applied to all lenses together, does the design's performance deteriorate too much?

We defined a dedicated single configuration Zemax file for the 410 GHz design at configuration 3 for the tolerancing calculation. This step was necessary because the original optical archive used the Zemax Multi-Configuration Editor to control the surfaces and coordinate systems associated with Prime-Cam's positions. Zemax does not permit tilt and decenter tolerances to be applied to surfaces that are controlled by multi-configuration operands, therefore configuration 3 was frozen into a standalone prescription before the alignment perturbations could be applied. We were interested in the response of the design to axial spacing errors, lateral decenter, and lens tilt. As such, the subsequent analysis should be interpreted as a rigid body alignment and assembly analysis, as opposed to a complete manufacturing tolerance budget.

\subsection{Alignment sensitivity analysis}
The aim of the inverse sensitivity analysis is to see how much degrees of freedom can be perturbed, to generate a target degradation in the merit function. We define an acceptable increase in the merit function, and then allow Zemax to determine the corresponding parameters to reach that degradation. The tolerancing calculation needed to focus on a criterion that responded directly to field dependent image quality. We constructed a tolerance merit function (MF) from 25 strehl, \textit{STRH} operands, one at each field point. Every field was set so the target Strehl ratio was equal to one, each with equal weighting. Lower MF values represent an average Strehl ratio closer to one. The nominal value of the MF was 0.2305. This provides the consistent criterion for comparing the nominal and perturbed systems. The lower the value, the more ideal the optical system. The mechanical degrees of freedom were chosen to represent likely uncertainties introduced from module fabrication and assembly. The two axial terms changed the separation between the first and second lens, and between the second and third lens. Each of the three lenses was also allowed to translate independently in \textit{x} and \textit{y}, and rotate about the two transverse axes. Decenter and tilt were applied to the pairs of surfaces bounding each lens, so that the perturbations would represent real rigid body motion of the lenses, as opposed to just changing one refracting surface. The tolerance operands are summarised in Table~\ref{tab:tolerance_budget}.

\begin{table}[H]
\centering
\small
\caption{Rigid-body alignment perturbations used in the 410 GHz configuration 3 Monte Carlo analysis. Decenter and tilt were applied to the surface pair bounding each lens so that the lens moved as a rigid element. All active terms were varied simultaneously, with paraxial focus compensation applied in each realisation.}
\begin{tabular}{l l l c}
\hline
Perturbation & Zemax operand & Element or spacing & Reference range \\
\hline
L1--L2 axial spacing & TTHI 47--48 & L1 to L2 & $\pm1.0$ mm \\
L2--L3 axial spacing & TTHI 52--53 & L2 to L3 & $\pm1.0$ mm \\
Lens decenter, $x$ and $y$ & TEDX/TEDY 44--45 & L1 & $\pm0.5$ mm \\
Lens decenter, $x$ and $y$ & TEDX/TEDY 49--50 & L2 & $\pm0.5$ mm \\
Lens decenter, $x$ and $y$ & TEDX/TEDY 58--59 & L3 & $\pm0.5$ mm \\
Lens tilt about $x$ and $y$ & TETX/TETY 44--45 & L1 & $\pm0.16^\circ$ \\
Lens tilt about $x$ and $y$ & TETX/TETY 49--50 & L2 & $\pm0.16^\circ$ \\
Lens tilt about $x$ and $y$ & TETX/TETY 58--59 & L3 & $\pm0.16^\circ$ \\
\hline
\end{tabular}
\label{tab:tolerance_budget}
\end{table}

An inverse sensitivity calculation was first used to examine each operand separately. Zemax varied one degree of freedom while allowing paraxial focus compensation, which is a selectable option in the tolerancing menu, and searched for a merit function increase of $\Delta$M = 0.01. This sensitivity criterion represents a 4.34\% change. The inverse calculation was used as a diagnostic rather than as a direct mechanical specification. Search intervals were expanded as operands hit the initial limit without producing the target degradation. This allowed weak sensitivities to be distinguished from results that were limited only by a narrow search interval. As such, large values appearing in the sensitivity output should not be read as proposed assembly tolerances. 

Table~\ref{tab:inverse_sensitivity} summarises the results of the inverse sensitivity analysis. The two axial spacing operands produced negligible changes in the tolerance merit function, even after the search interval was expanded. This means the design is less sensitive to lens spacing than rigid body angular and lateral alignment. The decenter and tilt alignment responses are more complicated. Several perturbations reached the target degradation, but the positive and negative directions were not generally symmetric. For example, positive L1 y-tilt reached the target at \textit{$0.322^\circ$}, whereas the opposite direction reduced the MF over the search range. L2 x-tilt reached the target at \textit{$-0.719^\circ$}, while positive tilt again moved toward lower MF values. L3 x-tilt reached the target at \textit{$0.546^\circ$} and at a much larger negative angle of \textit{$-4.264^\circ$}. The same directional behaviour appeared in the decenter results. Positive L2 x- and y-decenter reached the target at 3.410 mm and 2.985 mm, respectively, while the corresponding negative directions reduced the MF. L3 y-decenter reached the target at -1.396 mm but not in the positive direction within the search interval. L3 x-decenter was more symmetric, reaching the target at approximately $\pm4.983$ mm. 

This asymmetry can be attributed to the asymmetry in the optical system. A perturbation in one direction can therefore move the system toward a lower value of the MF, while the opposite direction moves it away. Moreover, the design is not centered on the optimal design so perturbations in certain directions can be expected to produce better optical performance, and therefore a lower MF. The key takeaways are that axial spacing produced very little response, whereas tilt and decenter generated meaningful changes at mechanically relevant scales. Within the tilt terms, L1 y-tilt and positive L3 x-tilt were among the tighter recovered limits. Among the lateral terms, negative L3 y-decenter gave the smallest decenter limit.

The inverse sensitivity results were then used to inform the ranges set in the Monte Carlo analysis. Using every limit the analysis returned, simultaneously, would not be appropriate as those were obtained by varying one degree of freedom at a time. For the Monte Carlo simulation, a common reference range was therefore assigned to each perturbation. These were adapted from the tolerancing analysis used for the Prime-Cam 850 GHz module\cite{huberOpticalMechanicalDetector2024}. The following ranges were selected: $\left|\Delta z\right| \leq 1.0\ \mathrm{mm}$ for both lens-to-lens spacings,
$\left|\Delta x\right|$,
$\left|\Delta y\right|$
$\leq 0.5\ \mathrm{mm}$
for lens decenter, and $\left|\theta_x\right|$, $\left|\theta_y\right|$ $\leq 0.16^\circ$ for lens tilt, as described in Table~\ref{tab:tolerance_budget}. These values are used as reference perturbation ranges for the sake of testing the design's sensitivity to errors, and are not a final mechanical tolerance range.

\begin{table}[H]
\centering
\small
\caption{One-at-a-time inverse-sensitivity results for the 410 GHz configuration 3 model. Values give the perturbation at which the Strehl-based merit function increased by 0.010 from its nominal value of 0.230462. ``Not reached'' indicates that the target degradation was not obtained within the expanded search interval; such entries should not be interpreted as mechanical specifications.}
\begin{tabular}{l c c l}
\hline
Perturbation & Negative direction & Positive direction & Interpretation \\
\hline
L1--L2 spacing & Not reached & Not reached & Negligible response in search range \\
L2--L3 spacing & Not reached & Not reached & Negligible response in search range \\
L1 $x$-decenter & Not reached & Not reached & Target not reached at $\pm10$ mm \\
L1 $y$-decenter & Not reached & Not reached & Target not reached at $\pm10$ mm \\
L1 $x$-tilt & Not reached & Not reached & Merit function decreased in both directions \\
L1 $y$-tilt & Not reached & $+0.322^\circ$ & Strong directional asymmetry \\
L2 $x$-decenter & Not reached & $+3.410$ mm & Strong directional asymmetry \\
L2 $y$-decenter & Not reached & $+2.985$ mm & Strong directional asymmetry \\
L2 $x$-tilt & $-0.719^\circ$ & Not reached & Strong directional asymmetry \\
L2 $y$-tilt & Not reached & $>+5^\circ$ & Approached but did not reach target \\
L3 $x$-decenter & $-4.983$ mm & $+4.983$ mm & Approximately symmetric \\
L3 $y$-decenter & $-1.396$ mm & Not reached & Tightest finite decenter result \\
L3 $x$-tilt & $-4.264^\circ$ & $+0.546^\circ$ & Strong directional asymmetry \\
L3 $y$-tilt & $-1.620^\circ$ & $+1.620^\circ$ & Approximately symmetric \\
\hline
\end{tabular}
\label{tab:inverse_sensitivity}
\end{table}

\subsection{Monte Carlo tolerance analysis}
The inverse sensitivity analysis isolates individual degrees of freedom, but an assembled module will not contain only one alignment error at a time. Lens spacing, lateral centering, and angular alignment will all be perturbed from their nominal values to some degree. The combined response will naturally differ from the sum of the individual perturbations. A Monte Carlo analysis was used to examine this combined behaviour. All 14 variables of interest were perturbed simultaneously, within the ranges defined in Table~\ref{tab:tolerance_budget}: two lens spacing terms, six decenter terms, and six tilt terms. Zemax generated 1000 random trials under the adopted tolerance model. In each trial, paraxial focus compensation was applied. This allowed a small image plane refocus to remove the purely defocus-like component of the perturbation. Without this compensator, an axial shift that would have been otherwise recoverable could dominate the merit function response and obscure the aberrations. 

\begin{figure}[H]
    \centering
    \includegraphics[width=0.8\linewidth]{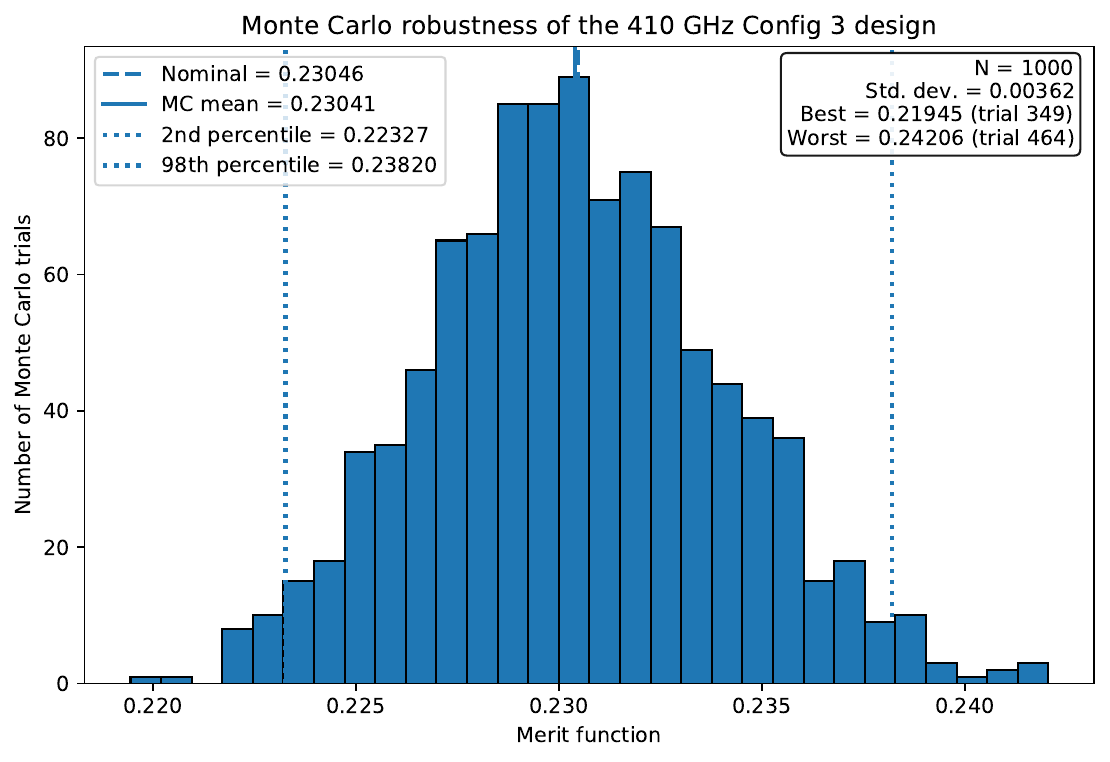}
    \caption{Distribution of the Strehl based tolerance merit function for 1000 Monte Carlo trials of the 410 GHz configuration 3 design. In each trial, the two lens spacings and the x- and y- decenter and tilt of all three lenses were varied simultaneously within the reference ranges in Table~\ref{tab:tolerance_budget}: paraxial focus compensation was applied to each trial. The distribution remains centred close to the nominal value, with mean 0.230405 and standard deviation 0.003625. The 2nd and 98th percentiles are labelled. Five of 1000 trials exceeded the inverse sensitivity target, and one trial exceeded a 5\% increase relative to nominal.}
    \label{fig:monte_carlo_histogram}
\end{figure}

Figure~\ref{fig:monte_carlo_histogram} shows the resulting distribution of the tolerance merit function. The nominal value is 0.2305, while the Monte Carlo mean is 0.2304. The distribution is therefore effectively centered on the nominal design. The slight offset has no practical significance and should not be interpreted as evidence that random alignment errors improve the optical system. The standard deviation is approximately 1.57\% of the nominal merit function value. The median, 0.2302, also lies close to nominal. The lowest merit function value was 0.2194 in trial 349. The highest was 0.2421 in trial 464. The worst trial therefore only increased the merit function by approximately 5.03\%. The upper tail is also small, with only 5 of the 1000 trials exceeding the merit function error of 0.01. The distribution is tightly concentrated around the nominal value, and unimodal. This result also helps to understand that the combined perturbations do not generally accumulate into a large degradation. Some combinations worsen the performance, while some partially compensate. This analysis establishes that the three lens design is not fragile to the tested placement errors. Within a tolerance model of $\pm$1.0 mm lens spacing, $\pm$0.5 mm decenter, and $\pm0.16^\circ$ tilt, the average Strehl merit function remains close to nominal in nearly all simulated designs.

From an engineering perspective, the inverse increment sensitivity analysis and Monte Carlo stages provide complementary guidance on proceeding. The inverse analysis indicates that mechanical effort is better spent controlling lens centering and angular alignment than pursuing unnecessarily tight lens-to-lens spacing. The combined analysis shows that within the reference alignment budget, a large degradation of optical performance is avoided. These results identify the degrees of freedom that require particular attention, however, they fail to establish achievable fabrication tolerances. These ranges should be revisited when the design reaches a more mature stage, to derive estimates better informed by the hardware. 

The perturbations in this study are applied directly to the lenses by varying their axial spacing, decenter, and tilt. In the instrument, however, these are not independent quantities - they arise from machining and assembly errors in the surrounding hardware, including welded shells, lens mounts, spacers, and support structures. As this hardware is positioned, small errors aggregate to determine the final position and orientation of each lens. The above optical analysis quantifies how sensitive the design is to those misalignments, but it does not determine how large the misalignments are expected to be in the manufactured module. To solve this, we developed an \textit{Excel} tool that propagates the machining tolerances through the assembly to predict the resulting lens spacing, decenter, and tilt. The 410 GHz instrument module welded shells have already been fabricated, and metrology measurements will be incorporated into this model to produce realistic predictions of the assembled lens positions. Those mechanically predicted lens perturbations can then be evaluated using the tolerancing methods presented here, allowing errors to be quantified in terms of their impact on optical performance. 

\section{Conclusion}
\label{sec:conclusion}
We evaluated whether the existing Prime-Cam 350 GHz three lens design can provide a practical baseline for a candidate 410 GHz instrument module. The design was tested without changing its lens surfaces, spacings, or receiver geometry, allowing the 410 GHz results to be interpreted as a direct test of the inherited design. The performance depends strongly on the effective position within Prime-Cam. Among the available configurations, we found configuration 3 to provide the strongest combined performance, at the reference telescope elevation considered. Across the field of view, it has a mean Strehl ratio of 0.838, mean ellipticity of 0.048, and mean $r_{80}$ of 1844 $\mu$m. Configuration 2 is substantially less favourable, with lower mean Strehl and larger field-to-field variation. Configurations 4 and 6 provide nearly identical intermediate performance. The direct frequency comparison shows that the shorter wavelength penalty remains moderate in the preferred configuration. For configuration 3, the mean Strehl ratio decreases from 0.879 at 350 GHz to 0.838 at 410 GHz, while the mean ellipticity only changes from 0.045 to 0.048, and the maximum ellipticity remains essentially unchanged. The design therefore loses some peak image quality at 410 GHz, as expected, but it does not develop a comparable field wide increase in beam asymmetry. The PSFs show why this conclusion cannot be based on Strehl ratio alone. The matched fields reveal compact beams, directionally extended beams, and structures with multiple lobes that are not ranked consistently by Strehl ratio, ellipticity, and encircled-energy radius. Configuration 3 is therefore preferred because it provides a better controlled distribution of beam properties over the field, not because it improves every metric at every individual field coordinate.

The alignment analysis provides an initial assessment of the design's sensitivity to specific reference errors. The inverse increment analysis calculations show that the Strehl based tolerance merit function is far less responsive to lens spacing errors than to lens decenter and tilt. Under the adopted tolerance model of $\pm$1.0 mm lens spacing, $\pm$0.5 mm decenter and $\pm$0.16$^{\circ}$ tilt, the 1000 trial Monte Carlo distribution remains tightly centred near the nominal merit function. Only five trials exceed the inverse sensitivity criterion. The design is therefore not unusually fragile to the tested combinations of alignment errors.

Within the scope of this study, the 350 GHz-derived three lens optical design remains a viable baseline for the Prime-Cam 410 GHz module.

\section{Acknowledgements}
The CCAT project, FYST and Prime-Cam instrument have been supported by generous contributions from the Fred M. Young, Jr. Charitable Trust, Cornell University, Duke University, and the Canada Foundation for Innovation and the Provinces of Ontario, Alberta, and British Columbia. The construction of the FYST telescope was supported by the Gro{\ss}ger{\"a}te-Programm of the German Science Foundation (Deutsche Forschungsgemeinschaft, DFG) under grant INST 216/733-1 FUGG, as well as funding from Universit{\"a}t zu K{\"o}ln, Universit{\"a}t Bonn, and the Max Planck Institut f{\"u}r Astrophysik, Garching. The construction of EoR-Spec is supported by NSF grant AST-2009767. The construction of the 350 GHz instrument module for Prime-Cam was supported by NSF grant AST-2117631. The construction of the 850 GHz instrument module for Prime-Cam is supported by CFI grants: 39656 and 46097 and Canadian provincial matching funds. The 410 GHz module is funded by the Canadian Foundation for Innovation, Innovation Fund 2025 Project 46097. The completion and deployment of the Prime-Cam instrument with the initial instrument modules is supported by a generous contribution from Alex Gerko, Founder and CEO of XTX Markets.

\bibliographystyle{spiebib}
\bibliography{references}

\begin{thebibliography}{10}

\bibitem{vavagiakisPrimeCamInstrumentOverview2026}
Vavagiakis, E.~M., Wang, Y., Lin, L.~T., Aravena, M., Austermann, J.~E., Bertoldi, F., Burgoyne, J., Butler, V., Chapman, S.~C., Choi, S.~K., Chung, D., Crites, A., Dev, A., Duell, C.~J., Fich, M., Fissel, L., Freundt, R.~G., Gazda, E., Huber, A.~I., Johnstone, D., Keller, B., Malachuk, P., Middleton, A., Moore, J., Niemack, M.~D., Nikola, T., Okada, Y., Patel, D.~A., Patel, T.~M., Riechers, D.~A., Scott, D., Stacey, G.~J., Vaughan, B.~J., Walker, S., Xie, R., and {{CCAT Collaboration}}, ``{{CCAT}}: The {{Prime-Cam}} instrument for the fred young submillimeter telescope---overview and status,'' Manuscript to be submitted to the Proceedings of SPIE (2026).

\bibitem{niemackDesignsLargeApertureTelescope2016}
Niemack, M.~D., ``Designs for a large-aperture telescope to map the {{CMB}} 10$\times$ faster,'' {\em Applied Optics}~{\bf 55},  1686--1694 (Mar. 2016).

\bibitem{parshleyOpticalDesignSixmeter2018}
Parshley, S.~C., Niemack, M.~D., Hills, R., Dicker, S.~R., D{\"u}nner, R., Erler, J., Gallardo, P.~A., Gudmundsson, J.~E., Herter, T., Koopman, B.~J., Limon, M., Matsuda, F.~T., Mauskopf, P., Riechers, D.~A., Stacey, G.~J., and Vavagiakis, E.~M., ``The optical design of the six-meter {{CCAT-prime}} and {{Simons Observatory}} telescopes,'' in [{\em Ground-Based and Airborne Telescopes VII}{\nolinebreak\hspace{0.1em}]},  Marshall, H.~K. and Spyromilio, J., eds.,  {\bf 10700},  1070041, SPIE, Austin, United States (July 2018).

\bibitem{vavagiakisPrimeCamFirstlightInstrument2018}
Vavagiakis, E.~M., Ahmed, Z., Ali, A., Basu, K., Battaglia, N., Bertoldi, F., Bond, R., Bustos, R., Chapman, S.~C., Chung, D., Coppi, G., Cothard, N.~F., Dicker, S., Duell, C.~J., Duff, S.~M., Erler, J., Fich, M., Galitzki, N., Gallardo, P.~A., Henderson, S.~W., Herter, T.~L., Hilton, G., Hubmayr, J., Irwin, K.~D., Koopman, B.~J., McMahon, J., Murray, N., Niemack, M.~D., Nikola, T., Nolta, M., {Orlowski-Scherer}, J.~L., Parshley, S.~C., Riechers, D.~A., Rossi, K., Scott, D., Sierra, C., {Silva-Feaver}, M., Simon, S.~M., Stacey, G.~J., Stevens, J.~R., Ullom, J.~N., Vissers, M.~R., Walker, S., Wollack, E.~J., Xu, Z., and Zhu, N., ``Prime-{{Cam}}: A first-light instrument for the {{CCAT-prime}} telescope,'' in [{\em Millimeter, {{Submillimeter}}, and {{Far-Infrared Detectors}} and {{Instrumentation}} for {{Astronomy IX}}}{\nolinebreak\hspace{0.1em}]},  Zmuidzinas, J. and Gao, J.-R., eds.,  1070804, SPIE, Austin, United States (July 2018).

\bibitem{huberCCATPrimeCamOptics2024}
Huber, Z.~B., Lin, L.~T., Vavagiakis, E.~M., Freundt, R.~G., Butler, V., Chapman, S.~C., Choi, S.~K., Crites, A.~T., Duell, C.~J., Gallardo, P.~A., Huber, A.~I., Keller, B., Middleton, A., Niemack, M.~D., Nikola, T., {Orlowski-Scherer}, J., Smith, E., Stacey, G., Walker, S., and Zou, B., ``{{CCAT}}: {{Prime-Cam}} optics overview and status update,'' in [{\em Millimeter, {{Submillimeter}}, and {{Far-Infrared Detectors}} and {{Instrumentation}} for {{Astronomy XII}}}{\nolinebreak\hspace{0.1em}]},  Zmuidzinas, J. and Gao, J.-R., eds.,  138, SPIE, Yokohama, Japan (Aug. 2024).

\bibitem{ccat-primecollaborationCCATprimeCollaborationScience2023}
{CCAT-Prime Collaboration}, Aravena, M., Austermann, J.~E., Basu, K., Battaglia, N., Beringue, B., Bertoldi, F., Bigiel, F., Bond, J.~R., Breysse, P.~C., Broughton, C., Bustos, R., Chapman, S.~C., Charmetant, M., Choi, S.~K., Chung, D.~T., Clark, S.~E., Cothard, N.~F., Crites, A.~T., Dev, A., Douglas, K., Duell, C.~J., D{\"u}nner, R., Ebina, H., Erler, J., Fich, M., Fissel, L.~M., Foreman, S., Freundt, R.~G., Gallardo, P.~A., Gao, J., Garc{\'i}a, P., Giovanelli, R., Golec, J.~E., Groppi, C.~E., Haynes, M.~P., Henke, D., Hensley, B., Herter, T., Higgins, R., Hlo{\v z}ek, R., Huber, A., Huber, Z., Hubmayr, J., Jackson, R., Johnstone, D., Karoumpis, C., Keating, L.~C., Komatsu, E., Li, Y., Magnelli, B., Matthews, B.~C., Mauskopf, P.~D., McMahon, J.~J., Meerburg, P.~D., Meyers, J., Muralidhara, V., Murray, N.~W., Niemack, M.~D., Nikola, T., Okada, Y., Puddu, R., Riechers, D.~A., Rosolowsky, E., Rossi, K., Rotermund, K., Roy, A., Sadavoy, S.~I., Schaaf, R., Schilke, P., Scott, D., Simon, R., Sinclair, A.~K.,
  Sivakoff, G.~R., Stacey, G.~J., Stutz, A.~M., Stutzki, J., Tahani, M., Thanjavur, K., Timmermann, R.~A., Ullom, J.~N., Engelen, A.~V., Vavagiakis, E.~M., Vissers, M.~R., Wheeler, J.~D., White, S. D.~M., Zhu, Y., and Zou, B., ``{{CCAT-prime Collaboration}}: {{Science Goals}} and {{Forecasts}} with {{Prime-Cam}} on the {{Fred Young Submillimeter Telescope}},'' {\em The Astrophysical Journal Supplement Series}~{\bf 264},  7 (Jan. 2023).

\bibitem{choiSensitivityPrimeCamInstrument2020}
Choi, S.~K., Austermann, J., Basu, K., Battaglia, N., Bertoldi, F., Chung, D.~T., Cothard, N.~F., Duff, S., Duell, C.~J., Gallardo, P.~A., Gao, J., Herter, T., Hubmayr, J., Niemack, M.~D., Nikola, T., Riechers, D., Rossi, K., Stacey, G.~J., Stevens, J.~R., Vavagiakis, E.~M., Vissers, M.~R., and Walker, S., ``Sensitivity of the {{Prime-Cam}} instrument on the {{CCAT-prime}} telescope,'' {\em Journal of Low Temperature Physics}~{\bf 199},  1089--1097 (May 2020).

\bibitem{Chapman2026}
Chapman, S.~C., Wheeler, J., Austermann, J., Beall, J.~A., Burgoyne, J., Devina, J., Henke, D., Huber, A., Hubmayr, J., Niemack, M., Patel, T., Sinclair, A., Vavagiakis, E., and Vissers, M.~R., ``The 410 {{GHz}} camera module for {{FYST}}: Design and testing of the {{MKID}} focal plane,'' Manuscript in preparation (2026).

\bibitem{kellerCCATDesignCharacterization2026}
Keller, B., Wheeler, J., Duell, C., Patel, D., Austermann, J.~E., Freundt, R.~G., Lin, L.~T., Middleton, A., Niemack, M.~D., Patel, T.~M., Vaskuri, A., Vavagiakis, E.~M., Walker, S., and Wang, Y., ``{{CCAT}}: Design and characterization of the 350 {{GHz}} instrument module,'' Manuscript in preparation (2026).

\bibitem{wyantBasicWavefrontAberration1992}
Wyant, J.~C. and Creath, K., ``Basic wavefront aberration theory for optical metrology,'' in [{\em Applied Optics and Optical Engineering}{\nolinebreak\hspace{0.1em}]},  Shannon, R.~R. and Wyant, J.~C., eds.,  {\bf 11},  1--53, Academic Press (1992).

\bibitem{dickerColdOpticalDesign2018}
Dicker, S.~R., Gallardo, P.~A., Gudmundsson, J.~E., Mauskopf, P.~D., Ali, A., Ashton, P.~C., Coppi, G., Devlin, M.~J., Galitzki, N., Ho, S.-P., Hill, C.~A., Hubmayr, J., Keating, B., Lee, A.~T., Limon, M., Matsuda, F., McMahon, J., Niemack, M.~D., {Orlowski-Scherer}, J.~L., Piccirillo, L., Salatino, M., Simon, S.~M., Staggs, S.~T., Thornton, R., Ullom, J.~N., Vavagiakis, E.~M., Wollack, E.~J., Xu, Z., and Zhu, N., ``Cold optical design for the large aperture {{Simons Observatory}} telescope,'' in [{\em Ground-Based and Airborne Telescopes VII}{\nolinebreak\hspace{0.1em}]},  Marshall, H.~K. and Spyromilio, J., eds.,  {\bf 10700},  107003E, SPIE, Austin, United States (July 2018).

\bibitem{gallardoSystematicUncertainties2018}
Gallardo, P.~A., Gudmundsson, J.~E., Koopman, B.~J., Matsuda, F.~T., Simon, S.~M., Ali, A., Bryan, S., Chinone, Y., Coppi, G., Cothard, N.~F., Devlin, M.~J., Dicker, S.~R., Fabbian, G., Galitzki, N., Hill, C.~A., Keating, B., Kusaka, A., Lashner, J., Lee, A.~T., Limon, M., Mauskopf, P.~D., McMahon, J., Nati, F., Niemack, M.~D., {Orlowski-Scherer}, J.~L., Parshley, S.~C., Puglisi, G., Reichardt, C.~L., Salatino, M., Staggs, S.~T., Suzuki, A., Vavagiakis, E.~M., Wollack, E.~J., Xu, Z., and Zhu, N., ``Systematic uncertainties in the {{Simons Observatory}}: Optical effects and sensitivity considerations,'' in [{\em Millimeter, Submillimeter, and Far-Infrared Detectors and Instrumentation for Astronomy IX}{\nolinebreak\hspace{0.1em}]},  Zmuidzinas, J. and Gao, J.-R., eds.,  {\bf 10708},  107083Y, SPIE, Austin, United States (Aug. 2018).

\bibitem{huberConstrainingTimeDependentParity2025}
Huber, Z.~B., {\em Constraining Time-Dependent Parity Violation with {{ACT}} and Developing Next-Generation Microwave Observatories}, phd dissertation, Cornell University, Ithaca, New York, United States (Aug. 2025).

\bibitem{huberCCATprimeOpticalCryogenic2022}
Huber, A.~I., Chapman, S.~C., Sinclair, A.~K., Spencer, L.~D., Austermann, J.~E., Choi, S.~K., Devina, J., Gallardo, P.~A., Henke, D., Huber, Z.~B., Keller, B., Li, Y., Lin, L.~T., Niemack, M.~D., Rossi, K.~M., Vavagiakis, E.~M., and Wheeler, J.~D., ``{{CCAT-prime}}: Optical and cryogenic design of the 850~{GHz} module for {{Prime-Cam}},'' in [{\em Millimeter, Submillimeter, and Far-Infrared Detectors and Instrumentation for Astronomy XI}{\nolinebreak\hspace{0.1em}]},  Zmuidzinas, J. and Gao, J.-R., eds.,  {\bf 12190},  121901D, SPIE (2022).

\bibitem{vavagiakisCCATprimeDesignModCam2022}
Vavagiakis, E.~M., Duell, C.~J., Austermann, J., Beall, J., Bhandarkar, T., Chapman, S.~C., Choi, S.~K., Coppi, G., Dicker, S., Devlin, M., Freundt, R.~G., Gao, J., Groppi, C., Herter, T.~L., Huber, Z.~B., Hubmayr, J., Johnstone, D., Keller, B., Kofman, A.~M., Li, Y., Mauskopf, P., McMahon, J., Moore, J., Murphy, C.~C., Niemack, M.~D., Nikola, T., {Orlowski-Scherer}, J., Rossi, K.~M., Sinclair, A.~K., Stacey, G.~J., Ullom, J., Vissers, M., Wheeler, J., Xu, Z., Zhu, N., and Zou, B., ``{{CCAT-prime}}: Design of the {{Mod-Cam}} receiver and 280 {{GHz MKID}} instrument module,'' in [{\em Millimeter, {{Submillimeter}}, and {{Far-Infrared Detectors}} and {{Instrumentation}} for {{Astronomy XI}}}{\nolinebreak\hspace{0.1em}]},  Zmuidzinas, J. and Gao, J.-R., eds.,  {\bf 12190},  1219004, SPIE (2022).

\bibitem{huberOpticalMechanicalDetector2024}
Huber, A.~I., {\em Optical, Mechanical, and Detector Developments for the Prime-Cam 850 GHz Module}, phd dissertation, University of Victoria, Victoria, British Columbia, Canada (2024).

\bibitem{chapmanCCATprime850GHzCamera2022}
Chapman, S.~C., Huber, A.~I., Sinclair, A.~K., Wheeler, J.~D., Austermann, J.~E., Beall, J., Burgoyne, J., Choi, S.~K., Crites, A., Duell, C.~J., Devina, J., Gao, J., Fich, M., Henke, D., Herter, T., Johnstone, D., Knee, L. B.~G., Niemack, M.~D., Rossi, K.~M., Stacey, G.~J., Tsuchitori, J., Ullom, J., Van~Lanen, J., Vavagiakis, E.~M., Vissers, M.~R., and {{CCAT-prime Collaboration}}, ``{{CCAT-prime}}: The 850 {{GHz}} camera for {{Prime-Cam}} on {{FYST}},'' in [{\em Millimeter, {{Submillimeter}}, and {{Far-Infrared Detectors}} and {{Instrumentation}} for {{Astronomy XI}}}{\nolinebreak\hspace{0.1em}]},  Zmuidzinas, J. and Gao, J.-R., eds.,  {\bf 12190},  1219005, SPIE, Montréal, Canada (Aug. 2022).

\bibitem{freundtStatusUpdateEoRSpec2024}
Freundt, R.~G., Li, Y., Henke, D., Austermann, J., Burgoyne, J.~R., Chapman, S.~C., Choi, S.~K., Duell, C.~J., Huber, Z.~B., Niemack, M.~D., Nikola, T., Lin, L.~T., Riechers, D.~A., Stacey, G.~J., Vaskuri, A.~K., Vavagiakis, E.~M., Wheeler, J., and Zou, B., ``{{CCAT}}: A status update on the {{EoR-Spec}} instrument module for {{Prime-Cam}},'' in [{\em Millimeter, {{Submillimeter}}, and {{Far-Infrared Detectors}} and {{Instrumentation}} for {{Astronomy XII}}}{\nolinebreak\hspace{0.1em}]},  Zmuidzinas, J. and Gao, J.-R., eds.,  {\bf 13102},  131020U, SPIE, Yokohama, Japan (Aug. 2024).

\bibitem{nikolaEpochReionizationSpectrometer2023}
Nikola, T., Choi, S.~K., Duell, C.~J., Freundt, R.~G., Huber, Z.~B., Li, Y., Malavalli, K., Niemack, M.~D., Rossi, K.~M., Stacey, G.~J., Vavagiakis, E.~M., Zou, B., Cothard, N.~F., Austermann, J., Wheeler, J.~D., Gao, J., Vissers, M.~R., Hubmayr, J., Beall, J., and Ullom, J., ``The epoch-of-reionization spectrometer, {{EoR-Spec}}: A spectrometer instrument module for {{Prime-Cam}} on {{FYST}},'' in [{\em Physics and Chemistry of Star Formation: The Dynamical {{ISM}} Across Time and Spatial Scales}{\nolinebreak\hspace{0.1em}]},  Ossenkopf-Okada, V., Schaaf, R., Breloy, I., and Stutzki, J., eds.,  352, Universit{\"a}ts- und Stadtbibliothek K{\"o}ln, K{\"o}ln, Germany (2023).

\bibitem{huberCCATprimeOpticalDesign2022}
Huber, Z.~B., Choi, S.~K., Duell, C.~J., Freundt, R.~G., Gallardo, P.~A., Keller, B., Li, Y., Lin, L.~T., Niemack, M.~D., Nikola, T., Riechers, D.~A., Stacey, G.~J., Vavagiakis, E.~M., and Zou, B., ``{{CCAT-prime}}: The optical design for the epoch of reionization spectrometer,'' in [{\em Millimeter, {{Submillimeter}}, and {{Far-Infrared Detectors}} and {{Instrumentation}} for {{Astronomy XI}}}{\nolinebreak\hspace{0.1em}]},  Zmuidzinas, J. and Gao, J.-R., eds.,  {\bf 12190},  1219019, SPIE (2022).

\bibitem{ansysHuygensPSF}
{Ansys, Inc.}, ``What is the difference between the {{FFT}} and {{Huygens PSF}}?.'' \url{https://optics.ansys.com/hc/en-us/articles/42661795535251-What-is-the-difference-between-the-FFT-and-Huygens-PSF} (2026).
\newblock Accessed 2026-07-21.

\bibitem{mahajanStrehlRatioPrimary1983}
Mahajan, V.~N., ``Strehl ratio for primary aberrations in terms of their aberration variance,'' {\em Journal of the Optical Society of America}~{\bf 73},  860--861 (June 1983).

\end{thebibliography}

\end{document}